\documentclass[reprint]{bioRxiv_mine}

\usepackage{enumitem} 
\usepackage{subcaption} 

\begin{document}

\title[Quantifying crowd noise in stadiums]{Towards a reproducible cross-venue method for quantifying crowd noise in stadiums}

\author{Alejandro Osses}
\altaffiliation{E-mail: \href{mailto:alejandro.osses@sorama.eu}{alejandro.osses@sorama.eu}, \href{mailto:info@sorama.eu}{info@sorama.eu}.\\ ORCID: \href{https://orcid.org/0000-0001-8028-4936}{0000-0001-8028-4936}.}
\author{Bente Ackermans}
\author{Helmer Nuijens}
\author{Rick Scholte}

\affiliation{Sorama BV, Eindhoven, the Netherlands}


\begin{abstract}
Public claims about the ``loudest stadium'' have been based on an instantaneous peak dB(A) reading measured at a single point, as popularised by the Guinness World Records ``loudest crowd roar'' category. The current record dates from 2014, where a maximum level of 142.2~dB(A) was registered. While compelling, those measurements lack standardisation, omitting relevant information such as the specific instrument that was used, the usage of time weighting, and the number of measurement positions that were tested. This lack of information does not allow a well-founded scientific comparison across sport venues. This study proposes a measurement framework in which spatially distributed acoustic measurement is the recommended route for a representative cross-venue comparison, while a single-anchor measurement can only serve as a minimum reporting baseline when distributed measurements are not feasible.
\end{abstract}

\maketitle

\section{Introduction}
\label{sec:introduction}

Sport venues with passionate supporters are renowned for their lively and engaging atmosphere, in which sound plays a central role. The acoustic environment in a stadium is not just a by-product of spectatorship, but it is rather an active contributor to the dynamics of the game~\cite{Siebein2024_LargeVenue}. In particular, players on the pitch are influenced ---both positively and negatively--- by the presence of crowd noise~\cite{Oldfield2026}. Elevated sound levels can enhance motivation and perceived home advantage, while at the same time interfere with communication, concentration, and decision making~\cite{Taylor2008, Barnard2011, Unkelback2010}. As such, the notion of a ``loud stadium'' is inherently tied to how sound is perceived by players during a game.

Beyond its impact on players, sound in stadiums has additional roles. It contributes to the overall spectator experience, reinforces collective engagement among supporters, and influences how events are perceived through broadcast media. These multiple roles make sound a key parameter in the evaluation of stadium atmosphere. Despite this, the quantification of ``crowd noise'' is typically focused on simplified metrics, often detached from the actual experience on the pitch.

Many crowd noise records exist and are often reported in press articles or compiled by organisations such as Guinness World Records (GWR), as summarised in Table~\ref{tab:01}.

\begin{table*}
\footnotesize
\caption{List of loudest stadiums with a reported  acoustic level. Most of these levels have been reported in A-weighted dB values, dB(A), however, when not explicitly mentioned, the unit is marked as ``dB.'' } \vspace{-8pt}
\scalebox{0.87}
{
\begin{tabular}{ccccccc} \hline\hline
Date & Home team & Venue & City & Sport & Level & Source \\ \hline
Sep 29, 2014 & Kansas City Chiefs & Arrowhead stadium & Kansas City, USA & American football & 142.2 dB(A) & \cite{GWR2014} \\
Dec 2, 2013  & Seattle Seahawks   & Lumen field       & Seattle, USA     & American football & 137.6 dB(A) & \cite{GWR2013_12} \\
Oct 13, 2013 & Kansas City Chiefs & Arrowhead stadium & Kansas City, USA & American football & 137.5 dB    & \cite{Chiefs2013}\\
Nov 8, 2023  & Tennessee          & Neyland stadium   & Knoxville, USA   & American football & 137.0 dB    & \cite{Saturday2023} \\
Sep 15, 2013 & Seattle Seahawks   & Lumen field       & Seattle, USA     & American football & 136.6 dB    & \cite{GWR2013_09} \\
Sep 19, 1999 & University of Washington & Husky stadium & Seattle, USA   & American football & 133.6 dB    & \cite{SeattleTimes2022} \\
2007         & Beşiktaş           & Vodafone Park     & Istanbul, Turkey & Football          & 132.0 dB    & \cite{Everton2025_Facebook} \\
Mar 18, 2011 & Galatasaray        & Ali Sami Yen stadium & Istanbul, Turkey & Football       & 131.76 dB(A)   & \cite{GWR2011} \\ 
Not reported & Fenerbahçe         & Şükrü Saracoğlu stadium & Istanbul, Turkey & Football    & 131.0 dB    & \cite{Everton2025_Facebook} \\ 
Feb 13, 2017 & Kansas Jayhawks    & Allen Fieldhouse  & Lawrence, Kansas, USA & Basketball   & 130.4 dB(A) & \cite{GWR2017} \\
Nov 5, 2024  & Celtic             & Celtic Park       & Glasgow, Scotland & Football         & 129.0 dB    & \cite{Everton2025_Facebook} \\
Aug 24, 2025 & Everton FC         & Hill Dickinson stadium & Liverpool, UK & Football & 126.0 dB & \cite{Everton2025_Facebook} \\ 
Oct 15, 2022 & Tennessee          & Neyland stadium   & Knoxville, USA    & American football & 125.4 dB & \cite{KnoxNews2022} \\ \hline\hline
\end{tabular}
}
\label{tab:01}
\end{table*}

The current loudest venue was registered in 2014 at Arrowhead stadium during an American football match~\cite{GWR2014}. The record was obtained as a peak reading from a sound level meter providing a level of 142.2~dB(A). It is unclear which specific instrument was used, where it was located, and how 
 this reading compares to the history of levels across different measurement positions, matches, and seasons. These observations are also valid for the other venues that are listed in Table~\ref{tab:01}.

While peak values are straightforward to communicate, they are limited from a scientific perspective. In particular, such values do not ensure representativity or reproducibility across venues and events~\cite{Acentech2019}. This limitation is particularly relevant when the quantity of interest is the sound as experienced by players, rather than the maximum level at a single measurement location. Several limitations that are associated with current practices are:

\begin{itemize}[leftmargin=*]
	\item The use of a peak reading is not aligned with standard acoustic procedures because transient or other extraneous sounds are typically not representative of the sound sources under investigation \cite{ISO1996-1_2016, ISO1996-2_2017}. Actually, transient sounds ---as those that can be picked up by a peak reading--- often require a separate treatment \cite{ISO1996-3_2022}. 
	\item If instead of a peak reading, the estimate of maximum levels is based on fast-integrated sound pressure levels, the estimations will be lowered by at least 10~dB~\cite{Acentech2019}, a non-negligible difference. This difference is illustrated later in Figure~\ref{fig:02-SLM}.
    \item The measurement position seems to be chosen based on intuition rather than using any spatial or acoustic criterion. Employing an acoustic criterion could ensure, for example, that the measurements are collected in the stable acoustic far field.
	\item Because there is no spatial criterion to choose the measurement locations, there is no guarantee that the highest crowd levels are being registered.
\end{itemize}

In other words, the current practices can lead to non-representative results or even to invalid acoustic measurements. The goal of this study is to establish the foundations of a more rigorous measurement protocol, so that the reliability of future crowd-noise estimates is increased. The adoption of such a protocol will also allow to~(1) characterise the crowd behaviour during specific match actions, making a distinction of the supporters' behaviour during, e.g., a goal or a free throw \cite{Barnard2011}; (2) assess noise exposure metrics \cite{Barnard2015, VanRansbeeck2024} as typically evaluated in hearing conservation programs of noise at work \cite{ISO9612_2025}; (3) project the estimates to predict exterior sound pressure levels, outside the stadium  \cite[e.g.,][]{Barnard2011}.

In this study we introduce a method to assess crowd levels at stadiums using one or multiple acoustic measurement devices. The method can use single microphone devices, such as sound level meters, but can also use distributed microphone arrays, also known as acoustic cameras. The structure of this paper is as follows: We start by identifying the limitations of the current crowd noise estimate (Section~\ref{sec:02}) and defining the criteria that must be fulfilled to ensure its reliable and representative quantification (Section~\ref{sec:03-considerations}). Based on those requirements, we first define a representative stadium crowd level and then describe a measurement framework (Section~\ref{sec:04}) in which spatially distributed measurements are the preferred implementation for a cross-venue comparison. We focus on practical and technical aspects that can lead to an objective and scalable noise monitoring system in indoor, semi-open, and outdoor sport venues. Finally we emphasise the scope of the method and indicate the limitations of the current proposal with an eye on future extensions of the method. 

\section{Loudest crowd roar: Current practice}
\label{sec:02}

\begin{figure}[!t]
	\centering
	\begin{subfigure}{0.42\textwidth}
	    \centering
		\includegraphics[width=.68\textwidth]{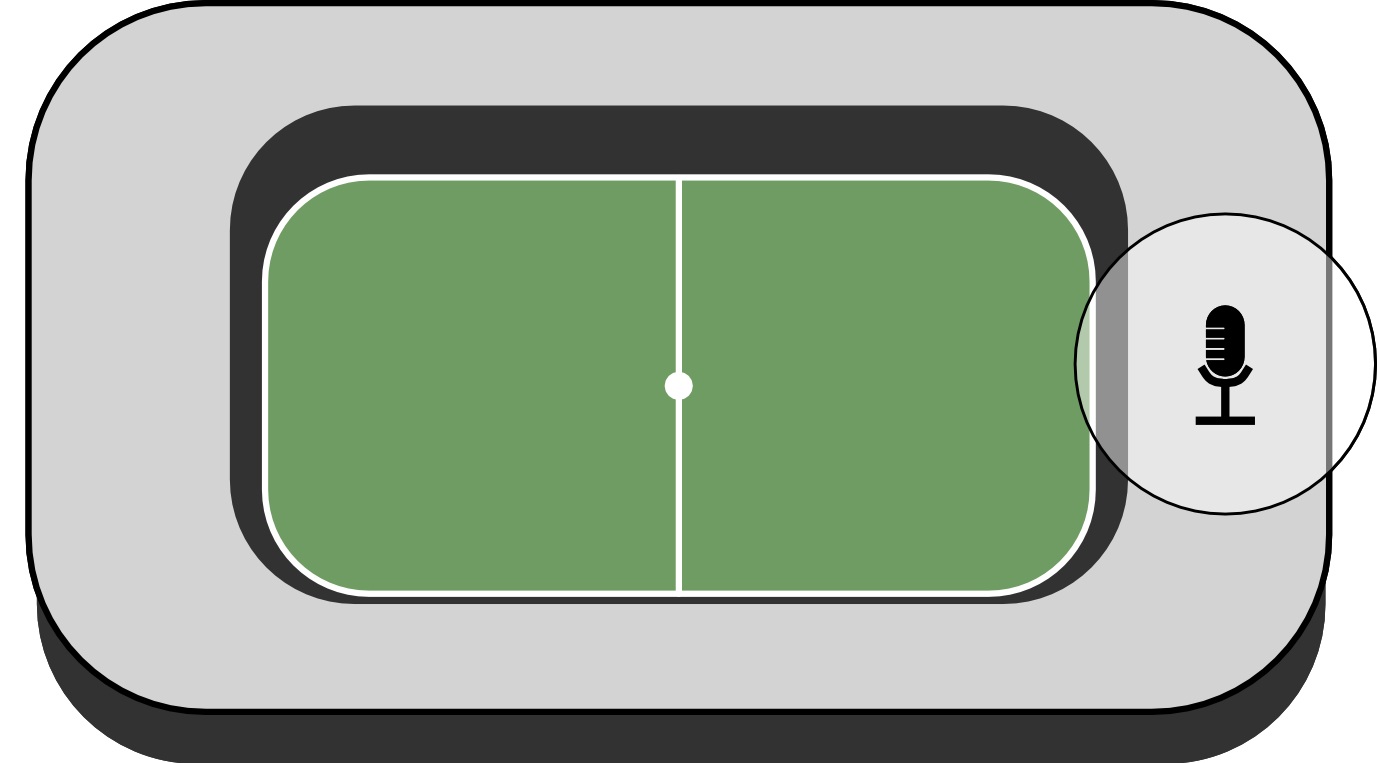}\\
        \vspace{-4pt}
		\caption{Current device positioning as used in the GWR procedure.}
	\end{subfigure}
	\begin{subfigure}{0.42\textwidth}
	    \centering
        \includegraphics[width=.68\textwidth]{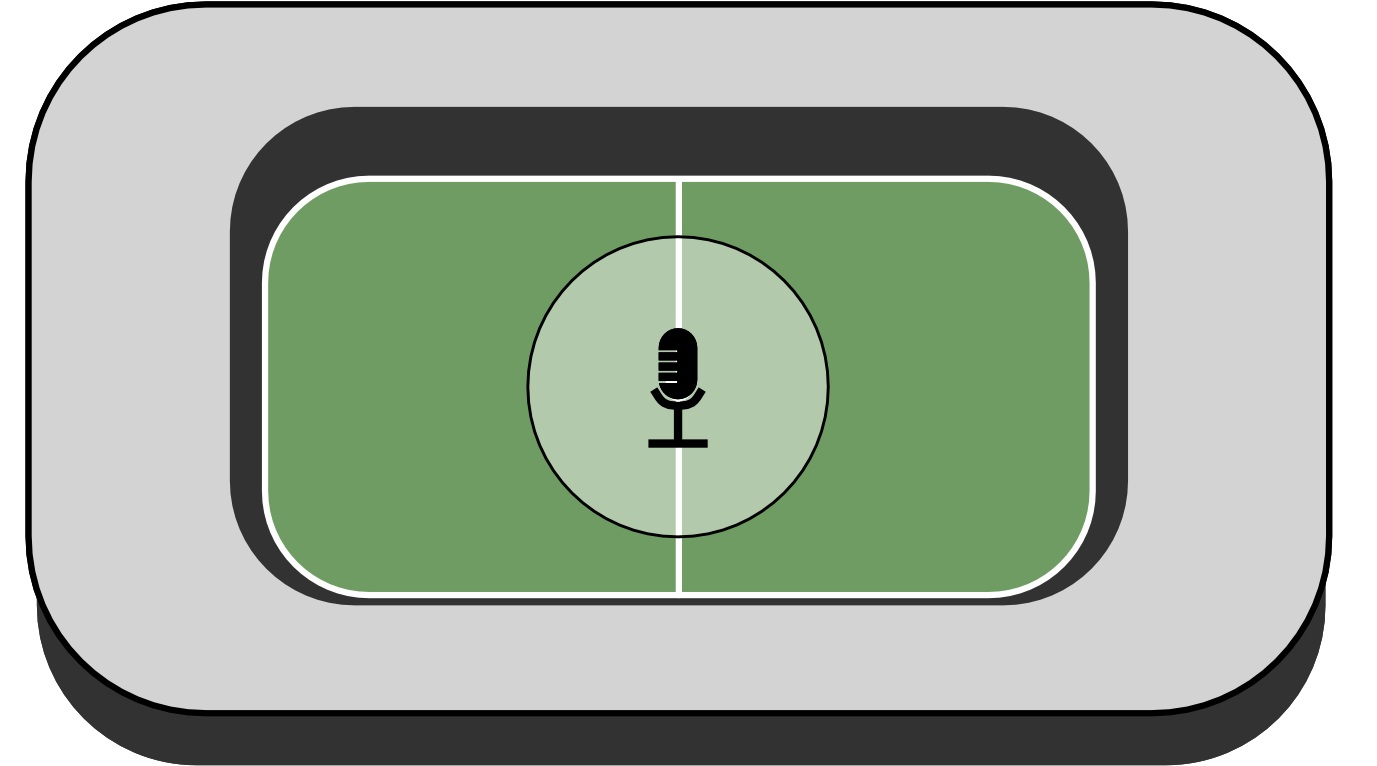}\\
        \vspace{-4pt}
	    \caption{One-device system that is pointed to the middle of the pitch. This arbitrary point will capture sounds coming from multiple audience sections.}
	\end{subfigure}	
    \vspace{-6pt}
	\caption{Generic stadium showing the pitch (in green) and audience areas (in grey), using one measurement device (microphone symbol). \textbf{Panel~(a)} reflects the device positioning as it has been so far used for crowd noise measurements, targeting a specific audience area, which is assumed to produce the highest crowd-noise levels. Two drawbacks of the current approach are: (1)~supporters in other sections do not get the possibility to become the loudest fans, because there is no measurement device in their neighbourhood, and (2)~the impact of the crowd onto the pitch, where the players are performing, is unknown. The one-device set-up shown in \textbf{Panel~(b)} partly copes with those limitations as it is pointed to the pitch which, by design, receives the level contribution from all the audience areas.}
	\label{fig:01-schematic}
\end{figure}

Despite the omission of specific technical details in the GWR procedure, in the category ``loudest crowd roar at a sports stadium,'' the method has been systematically used in the last decades \cite{GWR2014, GWR2017, GWR2013_09, GWR2013_12, SeattleTimes2022, Saturday2023, Chiefs2013}. The ``loudest crowd roar'' or, better said, the highest crowd noise level\footnote{Despite the fact that the term loudness or loudest crowd has been widely used, we will refrain as much as possible from its use. The reason is that loudness is actually the perceptual counterpart of sound pressure levels, expressed in Sones and not in dB as it is the case for sound levels~\cite{Zwicker1957, Osses2023Forum}. We will instead refer to ``crowd noise estimates'' or ``crowd levels.''}\nocite{Zwicker1957, Osses2023Forum} was claimed to be a peak dB(A) reading during a specific match moment. From press resources \cite[see, e.g., pictures from][]{GWR2014, GWR2013_09} can be seen that the metric has been based on point measurements collected with a Class~1 sound level meter Larson Davis 831, which was located very close to the crowd. A schematic stadium illustrating the positioning of such a single measurement device is depicted in Figure~\ref{fig:01-schematic}(a). 
 
The GWR approach disregards the impact of the crowd on the pitch and it does not give the possibility to supporters in other audience areas to become the loudest fans. These drawbacks  can be partly mitigated if the measurement device is targeted to the pitch, as depicted in Figure~\ref{fig:01-schematic}(b). We will come back to this discussion later, when multiple-device systems are introduced. 

\subsection{Why is a peak dB(A) reading ambiguous?}
\label{sec:dBA-ambiguous}

Peak levels are not typically used in environmental nor architectural acoustic standards. The main reason is that acoustic metrics have been developed to account for relevant aspects of human sound perception. One of these aspects is the difficulty of our ears to react to transient sounds. The ear is a ``sluggish'' organ that requires a time of about 200 ms to react to sudden level changes \cite[see][p.~64]{Moore2013}, resulting in sounds that are perceived as a slower and smoother time-varying envelope, despite the strongly-varying time information contained in a sound waveform. Historically, time weightings were adopted for tracking changes in level by a sound level meter due to early physical limitations of the old analog volume unit (VU) meters. Many years after, the fast integration time constant of 125~ms is well accepted as a correlate to the slower-varying perceived envelope and it is widely recommended in multiple international standards. According to ISO 1996-1 \cite{ISO1996-1_2016} that defines basic quantities used in environmental acoustics, two suitable metrics to characterise sound events are the fast-integrated maximum level $L$\textsubscript{AF,max} and continuous equivalent level $L$\textsubscript{Aeq}. The subindex ``A'' indicates the use of the A-weighting curve, whose use is another add-on to the level assessments that has been inspired by perceptual aspects, reflecting the ear's frequency-dependent sensitivity \cite{Houser2017}.

\begin{figure}[!b]
    \includegraphics[width=0.50\textwidth,trim=0 1.5cm 0 0,clip=true]{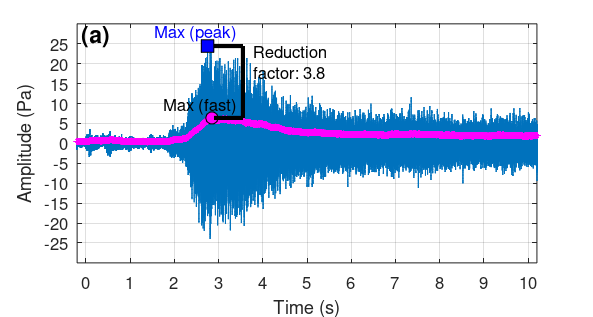}\vspace{-3pt}\\
	\includegraphics[width=0.50\textwidth,trim=0 0 0 0.8cm, clip=true]{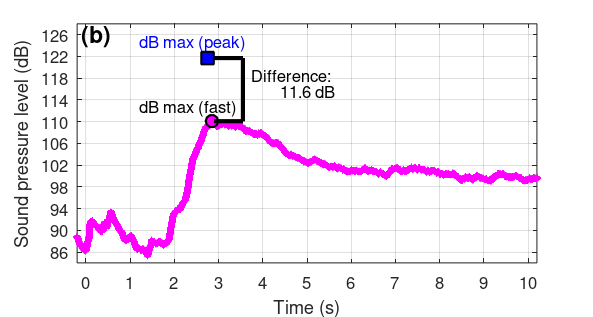}	
	\vspace{-20pt}
	\caption{Goal moment for a football match between PSV~Eindhoven and RKC~Waalwijk on 19 May 2024 from the Dutch Eredivisie. Ten seconds around the exact goal moment are drawn. \textbf{Panel~(a)} shows a waveform (in light blue) measured by an NTi XL2 Class~1 sound level meter which has a maximum of 24.4~Pascals or 121.7~dB (blue square marker). The pink trace shows the temporal envelope after the waveform was subjected to a fast time integration. The maximum of that curve is 6.4 Pascals or 110.1~dB (pink circle marker). \textbf{Panel~(b)} shows the envelope and the maximum estimates expressed in dB. The difference between the maximum level estimated from the peak reading and from the fast-integrated levels is 11.6~dB (reduction by a factor of 3.8).}
	\label{fig:02-SLM}	
\end{figure}

Considering all these arguments, a peak reading does not consider the time required to integrate a sound, and therefore is not a suitable measure of  perceived amplitudes. In fact, the exclusion of transient sounds is a common acoustic practice \cite[e.g.,][]{ISO1996-2_2017}. More generally, a peak reading can be seen as an estimator of a maximum value in a waveform. The waveform of a goal moment is shown in light blue in Figure~\ref{fig:02-SLM}(a) and its peak maximum is indicated by a blue square marker. The adoption of a fast time weighting will naturally smooth the waveform, converting it into a much slower time varying envelope (thick pink trace). The fast-integrated maximum in panel~(a) (pink circle marker) is 3.8 times smaller than the peak maximum reading. The maximum estimates and the envelope are expressed in dB in panel~(b), emphasising that for this example the peak reading overestimates the maximum level by 11.6~dB.
   
Another important aspect is that the measuring device will only be able to capture sounds that are under its microphone acoustic overload point. Acoustic overload points represent the maximum physical level of a measuring device before saturating, and are reported in dB(Z), i.e., before any time or frequency weighting is applied. Typical overload points are between 130 and 150~dB(Z). The reported overload point for the Larson Davis 831 sound level meter is 143~dB(Z)~\cite{LarsonDavis831}, which is suspiciously close to the reported 142.2~dB(A). This could mean that the GWR measurement might have been pushed down by the saturation point of the measurement device.

\subsection{Focus on the audience, the players, or both?}

So far, crowd noise estimates have been focused on sounds produced by the audience. This is probably the most straightforward way to estimate the supporters' strength and commitment to the team they support. We must not forget, however, that the supporters are primarily driven by the active players on the pitch and their actions \cite[e.g.,][]{Barnard2011}. If players can have such a direct impact on the supporters, why are the crowd noise estimates not focused on sound levels as projected onto the players?

A well-known case of the negative impact of crowd noise on a player was reported in 2017 during a football match between Beşiktaş and RB Leipzig at the Vodafone stadium (ranked 7\textsuperscript{th} in Table~\ref{tab:01}) valid for the UEFA Champions League~\cite{Valente2017_Werner}. During the game, the German player Timo Werner (RB Leipzig) requested the use of ear plugs and finally left the pitch after suffering from circulation problems as a consequence of the loud perception of the Beşiktaş' supporter roars. 

Focusing on sounds on the pitch seems to be a strategic choice because the reference measurement points will be confined in a smaller area with known dimensions, facilitating the comparison across stadiums. A schematic example of a straightforward point to spot on the pitch, the middle point of the venue, is illustrated in Figure~\ref{fig:01-schematic}(b). What needs to be investigated is whether crowd levels on the pitch will indeed be more representative of the overall atmosphere in the stadium compared to an evaluation purely focused on the audience. A data-driven discussion of this issue can help identify the most effective approach. 


\section{Improving the crowd noise estimation}
\label{sec:03-considerations}

\subsection{Frequencies of interest and their far field}

Environmental and room acoustics measurements implicitly assume that the data are collected satisfying a far field condition. The far field relates to the region in space where the underlying wavefronts of the sound source can be approximated as planar waves. The far field recommendation can be linked to the wavelength of the lowest frequency of interest.

\subsubsection{Target frequency range}

For stadiums and particularly focusing on sounds emitted by a crowd, we suggest the target frequency range between 100 and 8000~Hz \cite[see, e.g., Chapter~4 from][]{Bies2023}. This choice offers a pragmatic trade-off to capture:

\begin{itemize}[leftmargin=*]
    \item Fundamental frequency $f_0$ of loud voices (90-270~Hz): For male voices the average $f_0$ for vowels is 120 Hz, between 80 and 180~Hz, and for female voices is 200~Hz, between 150 and 260~Hz. A rise of about 10~Hz is expected for loud voices~\cite{Holmberg1988}.
    \item The effective range of vocal energy (300–3000~Hz): This range contains the formant components of vowels and voiced consonants~\cite{Ladefoged2006-Ch8}. 
    \item Consonant content (3000–8000~Hz) as found in fricative consonants (e.g., ``s,'' ``sh,'' and ``th'') and, to a lesser extent in plosive consonants (e.g., ``t,'' ``k,'' ``b,'' ``d,'' etc.)~\cite{Ladefoged2006-Ch8}. 
    \item Whistles and clapping transients as well as sharp cheering bursts (3000–8000~Hz)~\cite[e.g.,][]{Fu2025}.
	\item Chant detection (300–3000 Hz): This frequency range allows to keep all the relevant information to get intelligible speech~\cite[e.g.,][]{Greenwell2022}.
\end{itemize}

By focusing on measurements between 100 and 8000 Hz, very low frequency components that are unlikely related to crowds are excluded. Although this filtering seems very rough, slow frequency fluctuations ($<$20~Hz) that are relevant for the perception of sound reality \cite{Rosen1992, Osses2016a} are not affected. Those slow fluctuations are well retained by the temporal envelope in the 100-8000~Hz range \cite{Rosen1992}.

\subsubsection{Measurement distance: Far field criterion}

To ensure a far-field measurement under the classical acoustic assumptions, the lowest target frequency --–corresponding to the longest wavelength–-- must satisfy the far-field criterion. This wavelength represents the most restrictive condition for the validity of the far-field approximation.

In this study, we adopt the simplified relation:

\vspace{-8pt}
\begin{equation}
	r\textsubscript{critical}=2\cdot \lambda
	\label{eq:01}
\end{equation}

\begin{figure}[!t]
	\centering
	\begin{subfigure}{0.4\textwidth}
        \includegraphics[width=\textwidth,trim=0 15cm 0 0, clip=true]{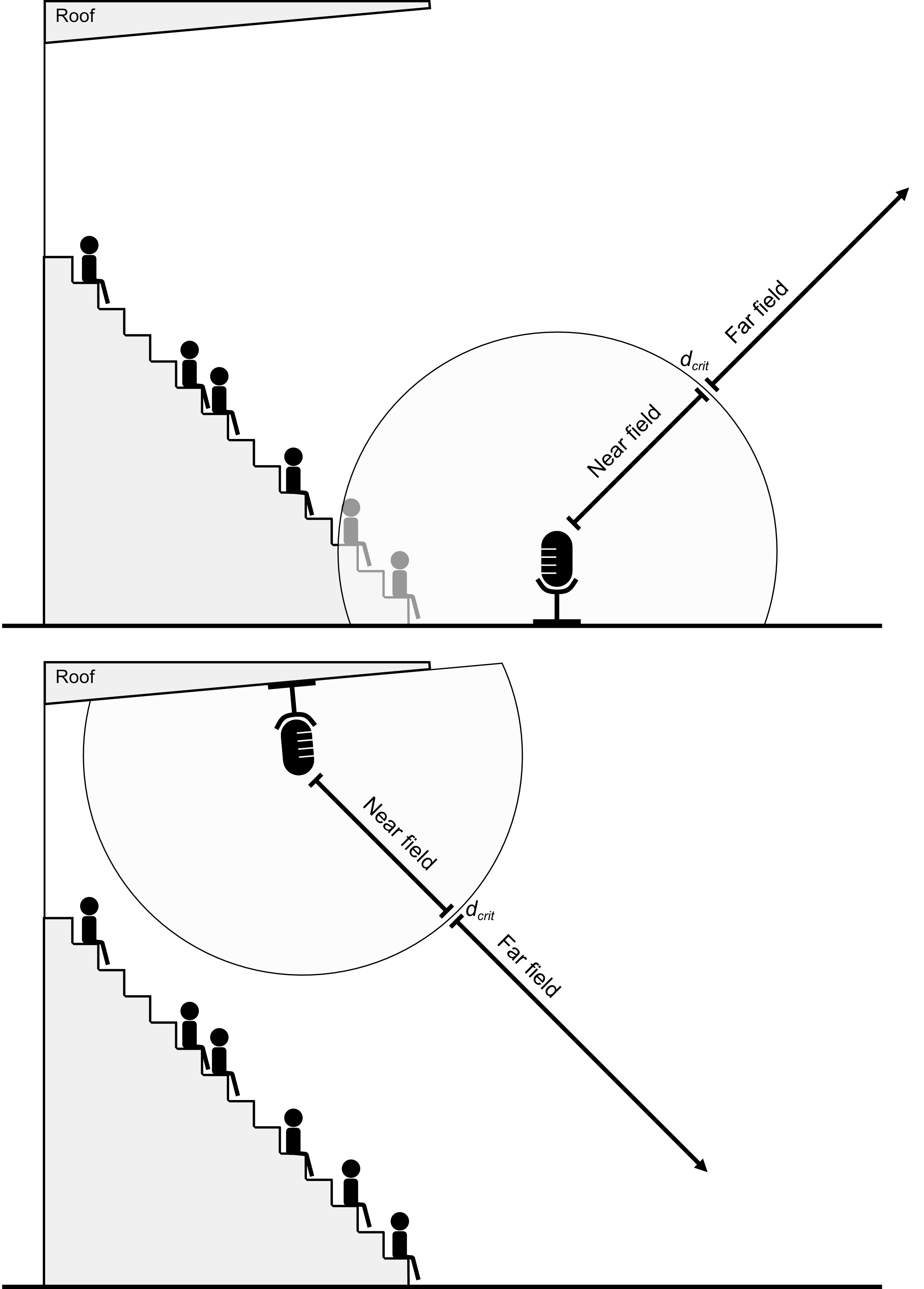}\\
        \vspace{-18pt}
        \caption{ }
   	\end{subfigure}
	\begin{subfigure}{0.4\textwidth}
        \includegraphics[width=\textwidth,trim=0 0 0 15cm, clip=true]{Fig3-Free-field-mod.jpg}\\
        \vspace{-18pt}        
        \caption{ }        
   	\end{subfigure}
	\vspace{-10pt}
	\caption{Position of a measuring device ---shown as a schematic microphone--- which is either \textbf{(a)} in front of the stands, as it could be done if a sound level meter is used for the measurements, or \textbf{(b)} on the roof, as a fixed installation, as it could be done when using a weatherproof continuous monitoring microphone or a microphone array. The critical distance $d\textsubscript{crit}$ is shown as the boundary between the near and far fields. The shaded areas represent the position in space where we discourage to obtain crowd level estimates. }
	\label{fig:03-free-field}
\end{figure}

\noindent where $r$\textsubscript{critical} is the minimum distance in meters between the sound source (i.e., the crowd) and the measurement position required to ensure the free-field conditions, and $\lambda$ is the wavelength in meters associated with the lowest frequency of interest.\footnote{Equation~(\ref{eq:01}) is independent of the size of the sound source under investigation. Such independence derives from the Fraunhofer distance where $r\textsubscript{critical} = \left(2\cdot D^2\right)/ \lambda$. Equation~(\ref{eq:01}) will be obtained when the characteristic dimension $D$ of the source is assumed to be acoustically small, conservatively using $D = \lambda$ (more generally with $D\leq\lambda$).}

A far-field measurement ensures that the measurement position lies outside both the reactive near-field and the transition region, where spherical spreading and stable radiation conditions can be assumed \cite{Kuttruff2007_Ch05}.

Considering that $ \lambda = c / f\textsubscript{low}$, with $c$ being the speed of sound equal to $\sim$340~m/s in air and $f\textsubscript{low}=100$~Hz being the lowest target frequency, and replacing these values in equation (\ref{eq:01}) result in a pragmatic minimum distance criterion $r\textsubscript{critical}$ of 6.8~m with respect to the audience area.\footnote{If for convenience a nominal $r\textsubscript{critical}$ of 8~m is used, the minimum frequency of interest $f$\textsubscript{low} is extended down to 85~Hz.} In larger stadium geometries or highly reflective environments, additional validation may be required. Figure~\ref{fig:03-free-field}(a) shows a schematic situation where the $r\textsubscript{critical}$ recommendation is not met for the two front-most seat rows, which are still in near field. For the same schematic stadium, Figure~\ref{fig:03-free-field}(b) shows a measurement position that meets with suggested far-field criterion.

\subsection{Requirements for a cross-venue stadium-noise metric}
\label{sec:03-2}

To ensure the comparability of a crowd-noise metric across stadiums we suggest some criteria related to: (1) the measurements to be conducted, and (2) the traceability of the instruments to be used. 

\subsubsection{Measurements}

\begin{itemize}[leftmargin=*]
	\item \textbf{Level readings (1)}: The measurements should be based on sound pressure level readings in dB(A).
	\item \textbf{Level readings (2)}: The levels should use a time integration equal to or greater than 125~ms, as defined for a fast time weighting \cite{IEC61672_1_2013}. A slow time weighting could also be adopted (time constant of 1 s), but longer time constants will generally be less sensitive to sudden crowd cheers. We suggest to adopt the standard fast time integration.
    \item \textbf{Location (1)}: The measurement points should be spatially meaningful: the exact position should be based on an objective criterion. The measurement point should dominantly capture human-generated sounds during the ``observation period.'' 
\item \textbf{Location (2)}: The distance between each measurement device and the audience area should be reported. The microphone(s) of the device should be~1.5 m or farther from any obstructing surface and pointed towards the audience \cite[e.g., see section~A.4 in][]{ISO3382-1_2009}. 
	\item \textbf{Data integrity}: The crowd noise estimate should be derived from sound levels that only contain actual crowd noise. Sounds from electronic amplification systems should be excluded from the crowd-level estimate where possible, or explicitly reported as separate contributions. Non-electronic supporter-generated instruments, such as drums, horns or vuvuzelas, should be reported separately or treated according to sport- and league-specific reporting rules.
\end{itemize}

\subsubsection{Traceability of the measurements}
\begin{itemize}[leftmargin=*]
	\item The measurement device should have a Class 1 level accuracy. We refer to ``Class 1 level accuracy'' devices to (1) sound level meters that comply with \citet{IEC61672_1_2013}, or (2) microphone arrays for which a frequency response measured at an angle of 0 degrees (in front of the device), complies with the acceptance limits defined in Table~3 from~\citet{IEC61672_1_2013}.  These limits should be met for the target frequency range of the analyses.
	\item The dynamic range of operation of the instrument should be reported. The target maximum levels should be below the acoustic overload point. If there is a risk of saturation during the measurements, this should be explicitly stated.
\end{itemize}

\subsubsection{Quantities to be reported}
\label{sec:quantities}

\begin{itemize}[leftmargin=*]
	\item \textbf{For specific sound events or short measurement intervals}: Levels should be logged using A weighting and a fast time integration. A suitable metric for reporting a frequency-weighted and time-integrated level is the maximum sound pressure level $L\textsubscript{AF,max}$. During the same measurement period, the continuous equivalent ``average'' level $L\textsubscript{Aeq,T}$ should be reported, including the duration $T$ of the observation. 
	\item \textbf{Match statistics (1)}: Data logging of continuous equivalent levels in periods of 1 s, 1 min and for other relevant event durations, e.g., $L\textsubscript{Aeq}$ values for each match quarter or half is recommended. For these longer term measurements it is desirable to report the percent exceedance (statistical) levels $L\textsubscript{AFNT}$, the acoustic percentiles, as defined in \citet{ISO1996-1_2016}. Recommended, $N=1, 5,$ or $10$ as metrics of longer term maximum levels during the match, and $N=90,$ or $95$ as an estimate of the minimum levels, which could serve as estimates of the background noise.
	\item \textbf{Match statistics (2)}: Relevant match statistics might be sport related. In that case, comparisons across stadiums can be difficult because each stadium constitutes a different source-path-receiver system, where the source, i.e., the supporters are very different from sport to sport. It may thus be relevant to consider the construction of within-sport databases, with the technical argument that the sound-source dependent transfer function of a stadium varies from sport to sport~\cite{Siebein2024_LargeVenue}. Multi-purpose sport venues may then have different crowd levels when used for different sports. 
\end{itemize}

\subsection{Field calibration and standards alignment}

\begin{itemize}[leftmargin=*]
	\item \textbf{Field calibration}: The measurement devices should be checked for field calibration before and after the events, so that drift logs are registered. Next to the field calibration, meteorological information (temperature and relative humidity) is desirable to be recorded.
	\item \textbf{Field validation of the system levels} (optional): Parallel measurements of the system for a cross check of levels with respect to a reference sound level meter: At a specific reference measurement position, simultaneous Class-1 sound level meter reading of $L\textsubscript{AF,max}$ and $L\textsubscript{Aeq,T}$ during outbursts. 
\end{itemize}

\subsection{What makes a good metric?}

Based on the previous considerations, a good crowd noise metric should:
\begin{itemize}[leftmargin=*]
	\item Be representative of the actual crowd noise levels at different stadium locations, including audience areas and/or players' locations on the pitch. The measurements should be conducted in the far field.
	\item Be robust to non-crowd noise. Transient sounds should not have a prominent effect on the metric. In addition, sounds that are not generated by the audience should be excluded from the analysis, as it is the case for sounds coming from the stadium public address system.
	\item Be non-intrusive to measure ideally using distributed devices deployed as a fixed system. The devices should be installed either on the roof, walls, or any other surface such that the safety in the stadium is never put at risk.
\end{itemize}

\subsection{Definition of a representative stadium crowd~level}
\label{sec:def-representative-level}

Based on the limitations of current practices and on the requirements outlined above, we define a representative stadium crowd level as a spatially representative estimate of the A-weighted sound pressure level generated by the audience during a live sport event, assessed under conditions that are relevant to the experience on the field of play, the pitch. This definition intentionally places the fan-generated contribution and the acoustic exposure of the active players at the centre of the assessment. In this context, sounds from public address systems, music playback, pyrotechnics, mechanical systems, or other non-crowd sources should be excluded from the crowd-level estimate, or explicitly reported as separate contributions when exclusion is not technically possible.
 
The stadium architecture and environmental conditions are part of the acoustic transmission path between the audience and the receiver positions. They may therefore influence the measured level and should be documented as part of the measurement context. For cross-venue comparison, it may be useful to distinguish between two complementary quantities: (1) a fan-generated crowd level, aimed at characterising the contribution of the supporters as the primary sound source, and (2) a venue-experienced crowd level, aimed at characterising the combined result of supporter behaviour, stadium geometry, enclosure, and atmospheric conditions at relevant receiver positions. In the present proposal, the priority is to define a transparent and reproducible reference quantity that enables fairer comparisons across venues than single-point peak measurements close to the crowd.

\section{Proposed measurement framework} 
\label{sec:04}

The definition of a representative stadium crowd level requires spatial representativeness. For that reason, the preferred implementation of the proposed framework is based on multiple measurement units distributed across the venue and, where relevant, directed towards the field of play. In situations where such a distributed system is not available, a single-anchor measurement can provide a minimum reporting baseline, provided that all requirements listed in Section~\ref{sec:03-2} are met and that the limitations of single-position measurements are explicitly reported.

\subsection{Single-anchor minimum baseline}
\label{sec:04-1}

A single-anchor metric should be interpreted as a minimum baseline rather than as the preferred method for representative stadium comparison, because the result remains dependent on the selected measurement position. To ensure spatial representativeness, the single-anchor metric is extended to a multi-anchor estimation in Section~\ref{sec:04-2}.

\subsubsection{Definition}

The measurement unit is used to compute a stadium crowd level (SCL\textsubscript{max}). The SCL\textsubscript{max} value is obtained as the $L\textsubscript{AF,max}$ value measured over a specific observation period that can be as long as the duration of the sport event or match. The observation period can also be related to halves or quarters of a match depending on the sport. If a shorter period is used, e.g., if a specific sound event is targeted, the length of the ``outburst window'' should be explicit and ideally motivated. If $i$ $L\textsubscript{AF,max}$ values have been logged, then the final metric SCL\textsubscript{max} can be obtained as:

\begin{equation}
	\mbox{SCL}\textsubscript{max} = \max{\left\{ \mbox{SCL}\textsubscript{max,i} \right\}}
	\label{eq:02}
\end{equation} 

\subsubsection{Why does this metric improve the current practice?}

The proposed procedure fixes time and frequency weightings to be applied to the sound pressure level readings, in alignment with current environmental and room acoustics guidelines. The procedure reduces the risk of (1)~inflating the level reading in case of the accidental measurement of a spurious sound source, and (2)~reporting sound levels that could have been measured violating the far field assumption. A strength of the proposed method is that it remains simple enough for its application using either sound level meters or distributed microphone arrays. The method is reproducible and transparent.

\subsection{Spatially-distributed estimator}
\label{sec:04-2}

For cross-venue comparison, the spatially distributed estimator is the preferred route because it reduces the dependence on a priori assumptions about the loudest audience section and provides a more representative description of the crowd contribution across the venue, in line with the definition from Section~\ref{sec:def-representative-level}.

\subsubsection{Definition}

The use of one anchor assumes that one specific device is representative of the stadium under study. The choice of an optimal anchor is therefore biased by prior knowledge of the loudest section in the stadium. To objectivise the choice of an ``optimal anchor,'' it is desirable to employ multiple measurement units distributed in the stadium. In such a setting, the sound events in the stadium are evaluated using multiple looks in a similar sense as adopted in signal detection theory \cite{Green1966}, having as many looks ---or independent observation points--- as units that are used. Hence, increasing the number of devices increases the probability of finding an overall maximum level of the stadium. Following the logic of the SCL\textsubscript{max} metric, the overall resulting metric can be obtained from the maximum of all maxima. To make a distinction in the notation, we introduce the prefix ``m'' to the name of the metric. If the SCL\textsubscript{max} value of device $k$ is named SCL\textsubscript{max,$k$}, then, the mSCL\textsubscript{max} metric is obtained as:

\vspace{-8pt}

\begin{equation}
	\mbox{mSCL}\textsubscript{max}=\max{\left\{ \mbox{SCL}\textsubscript{max,$k$} \right\}}
	\label{eq:multiUnit}
\end{equation}

Note that the mSCL estimation could use more elaborate ways of combining series of $L$\textsubscript{AF} values than the simple formulation in equation~(\ref{eq:multiUnit}), so that the mSCL metric accounts for the variability of levels between audience areas. We will return to this point in Section~\ref{sec:stats-instead-Lmax}.

\subsubsection{Schematic multi-unit stadiums}

\begin{figure}[!b]
	\centering
	\begin{subfigure}{0.36\textwidth}
	    \centering
	    \includegraphics[width=\textwidth]{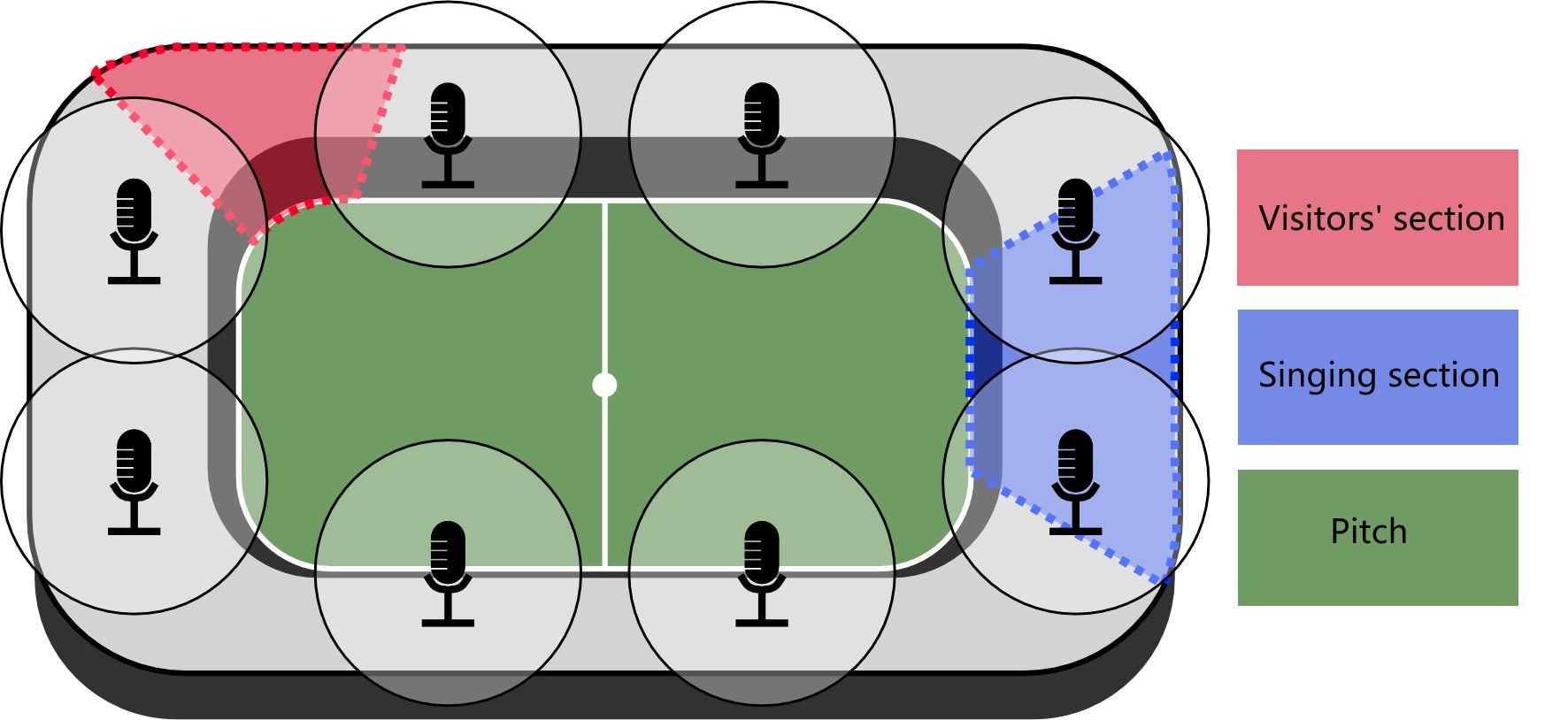}
	    \caption{Multi-unit system with $K$=8 devices.}
    \end{subfigure}
    \begin{subfigure}{0.36\textwidth}
        \vspace{-3pt}        
	    \centering
        \includegraphics[width=\textwidth]{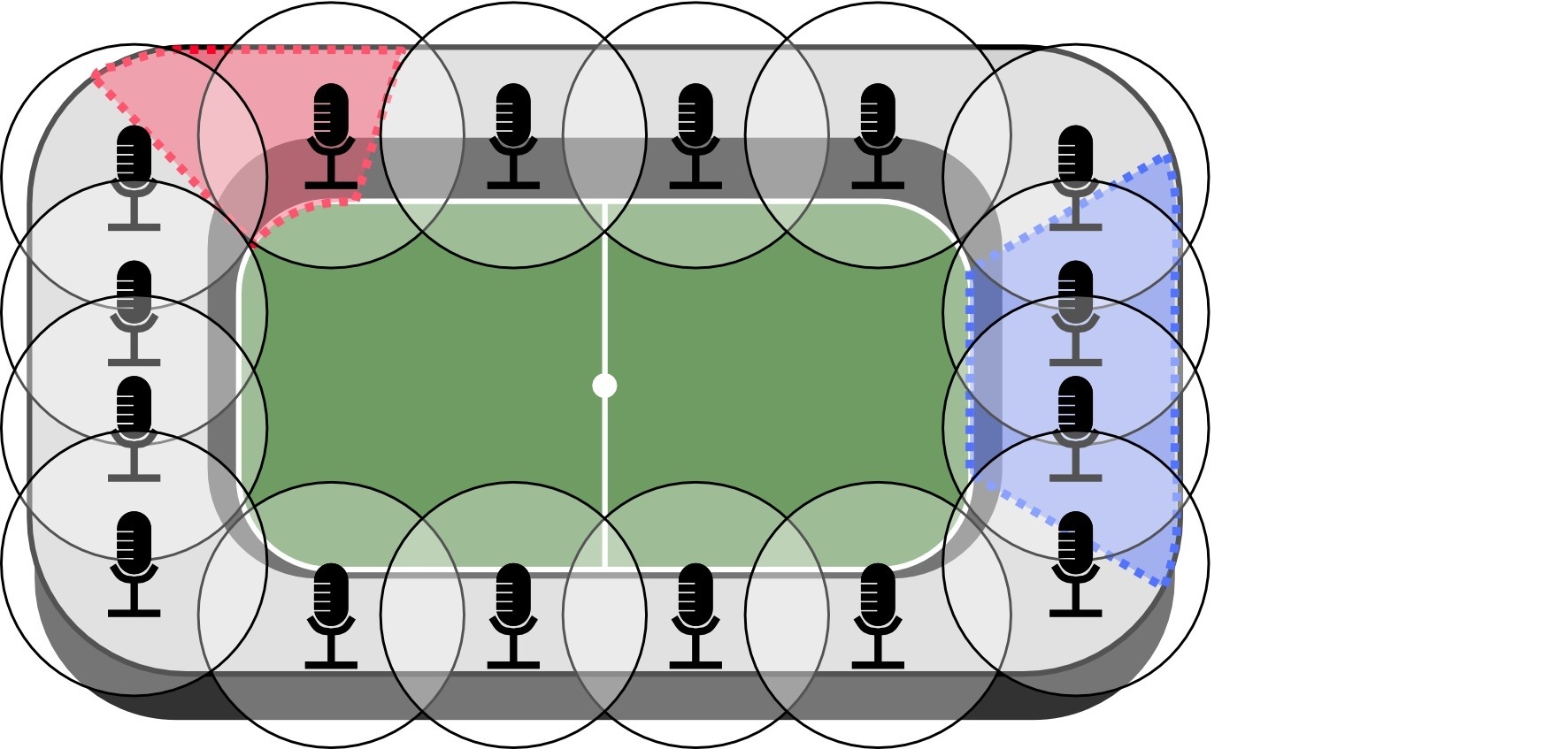}
	    \caption{Multi-unit system with a higher density of devices ($K$=16).}
    \end{subfigure}
    \begin{subfigure}{0.36\textwidth}
        \vspace{-3pt}    
	    \centering        
        \includegraphics[width=\textwidth]{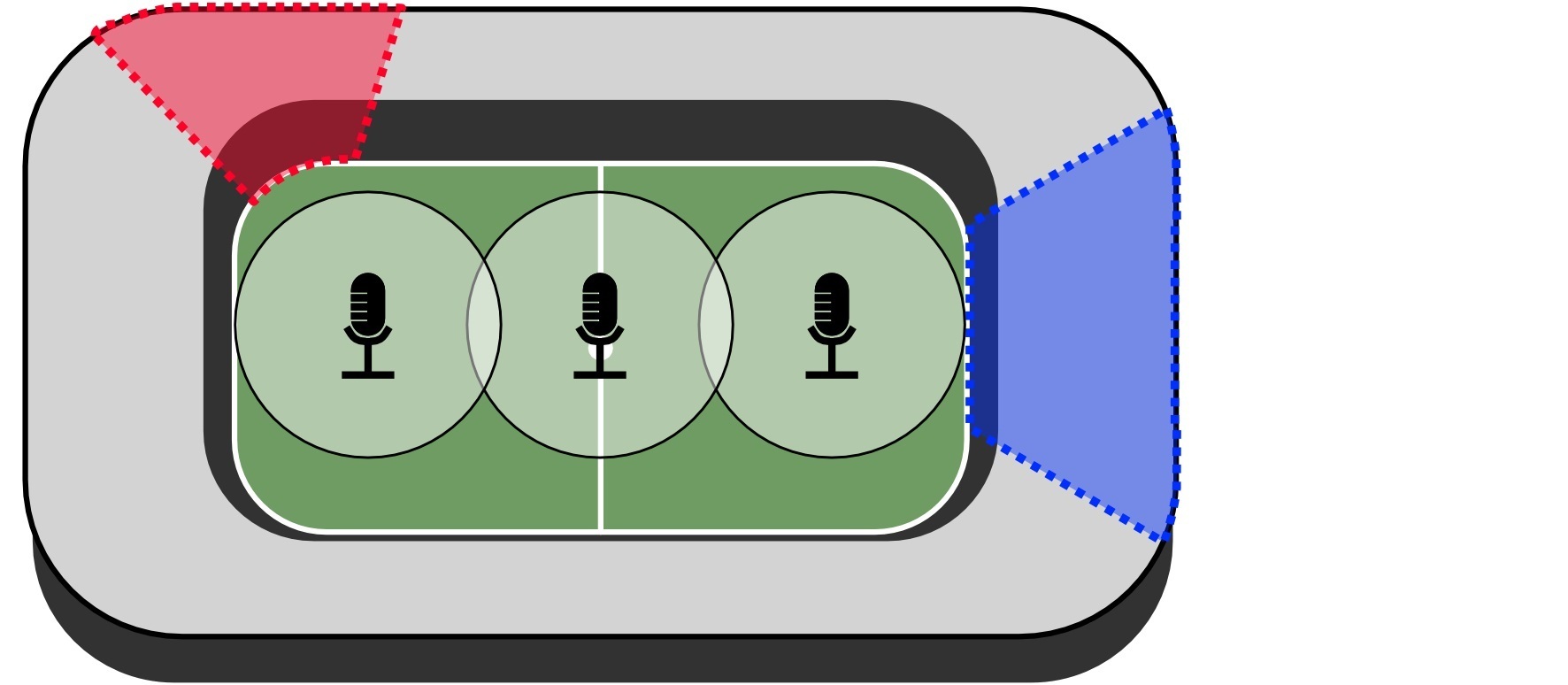}
	    \caption{Three-unit system distributed over the pitch area. This system can be combined with the system in panels (a) or (b), leading to a 9- or 19-device system, respectively.}
    \end{subfigure}
    \vspace{-6pt}
	\caption{Schematic of three multi-unit systems for a stereotypical stadium with seating sections around the pitch. The microphone symbols indicate distributed measurement devices. Two relevant target areas that could be related to singing and visitors' sections are highlighted in blue and red, respectively. Panels~(a) and (b) represent two systems that are increasingly more accurate than the one-device systems from Figure~\ref{fig:01-schematic}. Although these systems in panels (a) and (b) are still focused on the audience area only, they can be extended to cover the pitch if combined with the system from panel~(c).}
	\label{fig:04-schematic-with-sections}
\end{figure}

A schematic of two systems that can be installed in a stereotypical stadium is illustrated in Figure~\ref{fig:04-schematic-with-sections} with 8 and 16 devices. Assuming that the devices are hanging from the ceiling or dedicated trusses, grey circles indicate idealised acoustic fields of view, i.e., the area that is assumed to be reliably covered by the corresponding device. The seating sections are coloured blue for the singing section, red for the visitors section, and grey for the remaining sections. The microphone symbols indicate the measurement device locations. 

Two simple one-device systems were introduced in Figure~\ref{fig:01-schematic}. Figure~\ref{fig:01-schematic}(a) reflects the current GWR practice, targeted at one specific audience area. 

The system in Figure~\ref{fig:04-schematic-with-sections}(a) represents a system where arbitrarily two devices on each side of the stadium are placed, with a total of $K$=8 devices. Under this criterion, the schematic singing section is covered by two devices while the visitors section is still not directly covered by any of the other 6 devices. With the hypothesis that supporters from either the singing or visitors' sections can lead to the highest crowd noise levels, this system will still provide more advantage to the singing section. The visitors' cheering will however be much better covered, due to two devices that are close (but not close enough) to the visitors' area (in red).

The system in Figure~\ref{fig:04-schematic-with-sections}(b) represents a system with a fairer coverage of the whole stadium, where four devices are used along each side, with a total of $K$=16 devices. In this case, the visitors will be more accurately followed, giving them the possibility to drive the SCL estimate to them. Even if the visitors do not produce high levels, a comparison of the visitors' SCL and the SCL estimate for the singing section can lead to an interesting estimate of the stadium atmosphere, particularly for actions led by the home or the away team.

In order to extend the monitoring of crowd noises to other stadium locations, the systems in Figure~\ref{fig:04-schematic-with-sections} can be combined with one or more devices that cover the pitch area, as schematically drawn in Figure~\ref{fig:01-schematic}(b) and Figure~\ref{fig:04-schematic-with-sections}(c), respectively. 

\subsection{Device installation}

One of the challenges of a projection onto the pitch is related to the proper installation of devices that are sufficiently close to the pitch. If the stadium is open there might not be any mounting point above the pitch or the mounting possibilities may be very limited. This is in contrast to indoor stadiums where devices could be mounted on the roof or from trusses. Alternatives to ceiling installations [Figure~\ref{fig:01-schematic}(a)], are the installation of microphones alongside the pitch [Figure~\ref{fig:01-schematic}(b)] or even the use of microphones as wearable devices, similar in concept to referee-view broadcast systems in elite football~\cite{FifaRefereeView2026}. Under such installation constraints, accurate projections of a stadium level will require a more elaborate processing algorithm but, as pointed out earlier in this study, it could be relevant to study and keep track of historic levels projected to the field such that noise dosimetry estimates for players can be derived. 

\section{Discussion and perspectives for the future}

In the current study, we proposed a metric for the comparison of stadiums such that the levels of the crowd or ``loudness'' could be assessed. We presented a high-level description indicating relevant considerations related to the definition of the measurement locations and the acoustic metrics to be reported, such that future reports of crowd levels are based on valid acoustic measurements in contrast to the current records listed in Table~\ref{tab:01}. 

\subsection{Main recommendations}

The high-level description of the proposed measurement framework was focused on identifying weak choices, often omitted, in the measurement protocols that have been adopted in previous assessments of crowd noise levels.

For a fair cross-venue comparison, spatially distributed measurements should be regarded as the recommended approach to estimate crowd noise levels. A single-anchor measurement may be useful as a transparent minimum baseline, but it should not be interpreted as equivalent to a spatially representative stadium assessment.

More generally, we defined (1)~the minimum distance that is required to ensure a far field measurement, (2)~the recommendation of using fast-integrated A-weighted sound pressure levels, and (3)~the use of traceable instruments to collect the data. 

\subsection{Limitations and future work}

\subsubsection{Other statistics instead of the maximum levels}
\label{sec:stats-instead-Lmax}

The current SCL proposal is focused on estimations using maximum levels. While this choice aligns with the GWR procedure and improves it, it may still provide insufficient information for the appropriate characterisation of crowd noise at a stadium. As part of a revised version of the current SCL proposal, the following aspects need to be addressed and refined:

\begin{itemize}[leftmargin=*]
	\item Which specific level statistic is best suited to assess each individual anchor metric: next to the $L\textsubscript{AF,max}$ levels, $L\textsubscript{Aeq}$ values or other estimates may be more informative. These extra options may be sport dependent or even sound-event dependent (see Section~\ref{sec:quantities}).
	\item The combination of levels from multiple anchors using maximum levels [Equation (\ref{eq:multiUnit})] may not be sufficiently informative of, e.g., an overall maximum level as experienced by a player, or of the level contrasts between audience sections. 
\end{itemize}

The test cases under which the current multi-unit estimation will be representative and when a more elaborate combination of crowd levels between units will be required needs to be investigated.

\subsubsection{New test cases and crowd noise database}

Due to the high-level nature of the proposed method, no test cases are shared in this study. However, it is of crucial importance to have established an objective way of characterising crowd noise in stadiums first. The main motivation for this approach is the complex nature and diversity of sound events that take place during a sport event and the great variability that can be observed at different locations in the stadium. For an appropriate estimation of stadium crowd noise, we also postulate that the players' experience should be well captured, requiring to focus on crowd noise as projected onto the pitch. In addition, considering that many modern stadiums are multiuse and can even have retractable roofs~\cite{Siebein2024_LargeVenue}, it may be worth to treat multiuse stadiums as separate venues for different sports. 

To further advance the discussion of these aspects, we need to collect data and develop a database of sports venues with reliable estimates of crowd noise.

\subsubsection{Account for the exact acoustic environment}

We recommend to characterise a stadium based on far-field measurements. For this recommendation we linked the minimum distance for a reliable acoustic measurement to the lowest frequency of interest. Depending on the fulfilment of that criterion in addition to the use of multiple anchor devices, the measurement points and the audience can be very far apart. In this case, a distance compensation algorithm may be required. If this approach is needed, a physical correction that depends on distance~\cite{Ackermans2024, Osses2025}, temperature, relative humidity, or any other atmospheric or geometrical factor may be required. These aspects will be related to the type of acoustic field of the stadium ranging from open, semi-open, or close, and its specific dimensions. A procedure to obtain a reliable source-dependent transfer function may be needed in the future.

\subsubsection{Possibility of real-time noise mapping}

Although we encouraged the use of multiple devices, we did not emphasise the wide range of possibilities offered by the use of distributed microphone arrays, or acoustic cameras. In such a case, using the beamforming technique, local SCL\textsubscript{max} values can be projected to a group of seats, obtaining a noise map with a detailed grid. This is similar to the classical noise maps in acoustics \cite{Directive2002_49_EC} but more granular because each projected point will have its own level trace as a function of time. To convert such a map to a static map, any acoustic metric could be applied, e.g., by keeping the maximum obtained from each device, by averaging the $L$\textsubscript{AF} values, or by using other metrics such as percentiles applied to the time-varying estimates. These possibilities will be explored in future publications. Note that such noise maps can be employed to complement SCL\textsubscript{max} assessments in many different ways, that include building a database of historic levels for one stadium or across stadiums or for investigating the crowd behaviour for specific match events (e.g., goals, free throws, or free kicks).

\subsection{Final remarks}

In this study, we proposed a measurement framework for transparent and reproducible evaluation of crowd noise during sport events. The framework defines a representative stadium crowd level (SCL) as a spatially representative estimate of audience-generated A-weighted sound pressure levels, with spatially distributed measurement as the preferred route for cross-venue comparison. A single-anchor (SCL\textsubscript{max}) metric is retained as a transparent minimum reporting baseline when distributed measurement is not feasible, but should not be interpreted as equivalent to a spatially representative stadium assessment. The method is focused on how to use an acoustic measurement device that should minimally have one microphone, as is the case of the standard sound level meters, but it can actually use other devices. 

The main purpose of the current study is to raise awareness of the potential technical flaws of previous crowd noise measurements, to overcome these issues in future reports targeted to either scientific or general audiences. We also suggest that a fair crowd noise estimate should not only be representative of the levels in supporters' areas but should also account for the players' experience on the pitch. We invite the scientific community to discuss, scrutinise, and further develop measurement guidelines, so that future stadium level reports meet all the minimum acoustic standards. Such a synergy could also help to establish a procedure valid not only for mobile systems (point-based measurements) but also valid for installations of fixed  systems to which an uncertainty can be attributed. Such guidelines can help the future fair comparison of crowd noise levels across stadiums.

\bibliography{library-local}

\end{document}